\documentclass{IEEEcsmag}

\usepackage[colorlinks,urlcolor=blue,linkcolor=blue,citecolor=blue]{hyperref}

\usepackage{upmath}

\usepackage[capitalise]{cleveref}

\makeatletter
\usepackage{etoolbox}
\patchcmd\H@refstepcounter{\protected@edef}{\protected@xdef}{}{}
\makeatother

\usepackage{amsfonts}
\usepackage{tikz}

\jvol{XX}
\jnum{XX}
\paper{8}
\jmonth{May/June}
\jname{IT Professional}
\pubyear{2019}

\usepackage{listings}
\newcommand{\Test}[1]{\expandafter\hat#1}

\usepackage{soul}

\def\comment#1{\textsc{\color{blue}Comment: #1}}

\newif\ifdiff

\ifdiff
\def\added#1{{\color{green}#1}}
\def\removed#1{\textrm{\color{red}\st{#1}}}
\else
\def\added#1{#1}
\def\removed#1{}
\fi

\def\orcidID#1{\unskip$^{ORCID: [#1]}$}

\begin{document}


\sptitle{Department: Head}
\editor{Editor: Name, xxxx@email}

\title{Point Containment Queries on Ray Tracing Cores for AMR Flow Visualization}


\author{S. Zellmann\orcidID{0000-0003-2880-9090}}
\affil{University of Cologne and Bonn-Rhein-Sieg University of Applied Sciences, Institute of Visual Computing}


\author{D. Seifried\orcidID{0000-0002-0368-9160}}
\affil{University of Cologne, I. Physical Institute, Z\"ulpicher Str. 77, 50937 Cologne, Germany}

\author{N. Morrical\orcidID{0000-0002-2262-6974}}
\affil{NVIDIA \& SCI Institute}

\author{I. Wald\orcidID{0000-0003-0046-713X}}
\affil{NVIDIA}

\author{W. Usher\orcidID{0000-0001-5008-8280}}
\affil{Intel}

\author{J. A. P. Law-Smith\orcidID{0000-0001-8825-4790}}
\affil{Center for Astrophysics $|$ Harvard \& Smithsonian, Cambridge, MA 02138, USA}

\author{S. Walch-Gassner\orcidID{0000-0001-6941-7638}}
\affil{University of Cologne, I. Physical Institute, Z\"ulpicher Str. 77, 50937 Cologne, Germany}

\author{A. Hinkenjann\orcidID{0000-0002-8391-7652}}
\affil{Bonn-Rhein-Sieg University of Applied Sciences, Institute of Visual Computing}

\markboth{Department Head}{Paper title}



\begin{abstract}
Modern GPUs come with dedicated hardware to perform ray/triangle intersections and \added{bounding volume hierarchy} (BVH) traversal.
While the primary use case for this hardware is photorealistic 3D computer graphics, with careful algorithm design scientists can also use this 
special-purpose hardware to accelerate general-purpose computations such as point containment queries. 
This article explains the principles behind these techniques and their application to \removed{flowfield}\added{vector field} visualization of large simulation 
data using particle tracing.
\end{abstract}

\maketitle


\chapterinitial{Ray tracing} is a computer graphics algorithm that is based on 
geometric optics. While traditionally being focused on offline production 
rendering by the film industry, the recent addition of ray tracing cores, or 
RT cores, to graphics processing units (GPUs) has led to broader adoption of 
this technique for real-time applications. Ultimately, this newly gained popularity 
can be attributed to increased throughput in terms of the number of rays that can 
be traced through a given 3D geometry in a unit of time.

RT cores have the purpose of geometrically intersecting rays
that are defined by their 3D origin $o \in \mathbb{R}^3$ and an (often unit) 
direction vector $\vec{d} \in \mathbb{R}^3$ with 3D objects that are usually 
given by a parametric equation, solving for the distance $t$ of the point of 
intersection to the ray's origin. Sometimes it is also necessary to restrict 
the objects tested to some interval $\left[t_{min},t_{max}\right]$ along the 
ray, for example, when the ray is used to compute shadows where we are only 
interested in the objects between the origin and the light source. Therefore, 
most ray tracing APIs also store this interval along with the ray. Ray/object 
intersections are usually accelerated using search data structures, the most 
popular of which is arguably the bounding volume hierarchy (BVH). RT cores by 
NVIDIA, for example, have dedicated hardware for ray/triangle intersection and 
ray/BVH traversal.

In this regard, RT cores are essentially hardware units that accelerate tree traversal. 
Therefore, it is possible to use these units to accelerate general tree-traversal-based 
computations and not only graphics. The only requirement is that one can reformulate their 
algorithms to be mapped to the ray tracing hardware. For example, Zellmann et al.~\cite{zellmann:2020c} 
used RT cores to accelerate force-directed graph drawing, where the efficacy of spring forces 
decreases with distance. The computations involved can be accelerated using a search tree over 
the graph's vertex set.
The vertices are interpreted as particles, and the forces are computed 
using radius point containment queries.

Reformulating one's algorithm is, of course, not always possible or efficient. 
For example, the problem domain must be embeddable in $\mathbb{R}^3$, and in
general, the problem must be mappable to a ray tracing problem without introducing 
significant overhead. In the previous example of the radius point queries~\added{\cite{zellmann:2020c}}, the 
authors observed that a fixed radius point query over a set of particles 
could be reformulated as a ray-tracing problem by using an inverse mapping. 
The authors promoted the set of particles to a set of (generally overlapping) 
spheres, and the sphere of interest inside which the neighboring particles are 
gathered becomes a ray with length zero. 

Observing that the above motivating example of point queries over a set of particles 
draws its inspiration from physics---in fact, force-directed graph drawing can also be 
implemented as an n-body simulation~\cite{eades:1984}---it becomes evident that this 
overall principle might apply to all sorts of scientific fields such as fluid dynamics 
or flow visualization. The motivating example for this article is particle tracing in 
\removed{flowfields}\added{vector fields} based on a particular grid representation---namely adaptive mesh refinement 
topologies that are common in astrophysical or meteorological simulation codes.

\section{Point Containment Queries with NVIDIA RTX and OptiX}
\begin{figure}
\centering
\begin{tikzpicture}
\node[anchor=south west,inner sep=0] (image) at (0,0) {\includegraphics[width=0.99\linewidth]{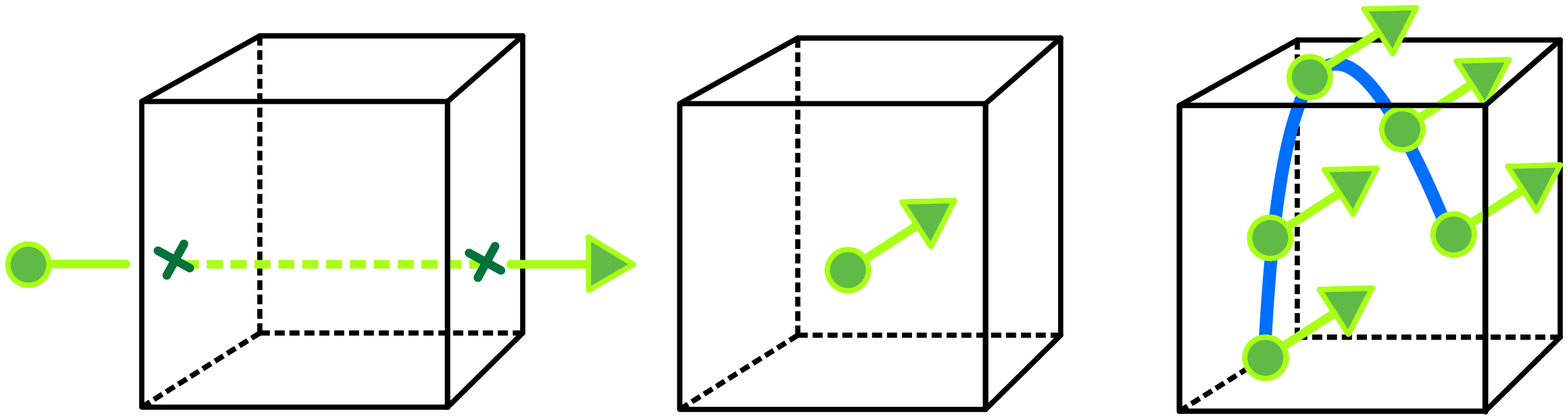}};
\node[anchor=south west,inner sep=0] at (1.2,-0.4) {(a)};
\node[anchor=south west,inner sep=0] at (3.7,-0.4) {(b)};
\node[anchor=south west,inner sep=0] at (6.2,-0.4) {(c)};
\node[anchor=south west,inner sep=0] at (0,0.3) {$t_{min}$};
\node[anchor=south west,inner sep=0] at (2.2,0.3) {$t_{max}$};
\node[anchor=south west,inner sep=0] at (3.9,0.30) {$\lim_{}$};
\node[anchor=south west,inner sep=0] at (3.6,0.1) {\footnotesize{$\Delta t\to0$}};
\end{tikzpicture}
\caption{\label{fig:pointqueries}
Ray-traced point containment queries. For simplicity, the integration domain here is deliberately
just a single box, but would typically be comprised of multiple overlapping
finite elements or grid cells. (a) Ordinary ray tracing only allows us to compute
the intersections with the elements' \emph{boundaries} in the interval of length $\Delta t = t_{max}$-$t_{min}$.
(b) If the length of that interval is zero the ray becomes a single point so we can find the \emph{overlapping} elements
at the ray's origin position. (c) This principle can be used to perform general
point containment queries to, for example, compute streamlines if
the integration domain represents a \removed{flowfield}\added{vector field}.
}
\end{figure}
Point containment queries with RT cores have been proposed by several researchers,
including the aforementioned force-directed graph drawing algorithm by
Zellmann et al.~\cite{zellmann:2020c}. Even more closely related to our approach is the
work by Morrical et al.~\cite{morrical:2019} who used point containment queries to evaluate
the particle density at arbitrary sampling positions $\mathbf{x} \in \mathbb{R}^3$
given an underlying unstructured grid representation with finite elements such
as tetrahedra or hexahedra.

By building a \removed{RTX} BVH over the set of finite elements \added{using one of the SDKs that support hardware ray tracing on NVIDIA GPUs (i.e., OptiX, DXR, or Vulkan)},
point containment queries can again be performed elegantly by just tracing a ray of length zero (see \cref{fig:pointqueries}) through the hierarchy\added{.}
\removed{where}\added{For that,} we just set $t_{min},t_{max}=0$ and the direction to some arbitrary non-zero-length value (for example $\vec{d} =
(1,1,1)$). \added{On NVIDIA GPUs that are labeled RTX, those
point containment queries are hardware-accelerated.}

With NVIDIA's RTX extensions, the proper way to implement point containment queries is through \emph{user geometry} representing the finite elements. We briefly review the steps involved in setting up this user geometry, and along the way
also introduce the necessary terminology. The description and terminology we use
roughly follows the OptiX API.\footnote{\url{https://raytracing-docs.nvidia.com/optix7/index.html}}

OptiX exposes entry points for the programmer to initiate and control ray
generation and ray/object interactions. The \emph{ray generation program}
is comparable in nature to a compute kernel as it is executed by the threads
of the programmable shading multiprocessors. In the
case of a program that performs point containment queries, the ray generation
program would be the place where the zero-length rays are generated at the appropriate
positions. In the case of particle tracing, the
ray generation program would hence also be used in the same way as one
would use a compute kernel to implement particle tracing,
by performing accesses to the computation domain
through an OptiX BVH.

The point containment query rays of length zero are traced against this BVH, which is built
using OptiX host API routines before the device program
starts execution, and whose handle is then passed over to the ray
generation program. OptiX supports both triangle BVHs, in which
case the whole ensuing traversal routine is executed in hardware, or
user geometry BVHs where the user specifies an \emph{intersection program}
that has access to the ray and that reports an intersection if the ray was found to
intersect the user geometry.
In the user geometry case, when the ray is traced, the hardware will perform many context switches between
hardware BVH traversal and software intersection programs. Those context switches come
at a performance penalty that gets worse the more often the context switches
are necessary.

Point containment queries require the
use of an intersection program and hence must be implemented using user geometry.
The intersection program will report an intersection if the ray origin is contained
inside the element. The list of elements that contain the point of
interest can be stored and updated in a thread-local data structure (often referred to as the
``per ray data'') and is available in both the \emph{closest hit program} that
is executed right after traversal finished, as well as in the ray generation
program where execution is returned to afterwards.

\section{Adaptive Mesh Refinement}
Simulation codes compute and output quantities such as density, velocity or
other scalar and vector fields for a three-dimensional domain that are assigned
to data/sample points. The spatial arrangement of those
data points defines the topology of the domain and is crucial to the
computational and memory efficiency of the simulation code. The simplest topologies
use structured grids with uniform cell sizes. Since memory bandwidth over the years has not
grown at the same rate that transistor density has, more efficient
topologies distribute the \emph{sample point budget} to 3D regions where the
entropy is relatively high. In particular, those topologies are usually either
the already motivated, generally unstructured finite element meshes,
or they fall into the category of adaptive mesh
refinement (AMR) where the topology is \emph{locally} structured and globally
connected using a hierarchy. \added{In practice, }AMR topologies come in many forms
\removed{, examples being
block-structured AMR, Octree AMR, or
Cartesian AMR.} \added{such as Octree or block-structured AMR, where the
major differences are in regards to branching factor and number of cells stored
at different hierarchy levels}.
\removed{AMR is not limited to
astrophysics or meteorology, but is also used in engineering. Hariharan for example used an AMR
solver to simulate vorticity surrounding rotorcraft (helicopters). Their paper is 
closely related to ours as its topic is AMR flow visualization, too. The authors
however do not have a separate processing phase to extract streamlines, because in
their case the flow field forms a clearly discernible isosurface that, when visualized,
conveys all the relevant information.}

\begin{figure}
\centering
\begin{tikzpicture}
\node[anchor=south west,inner sep=0] (image) at (0,0) {\includegraphics[width=0.99\linewidth]{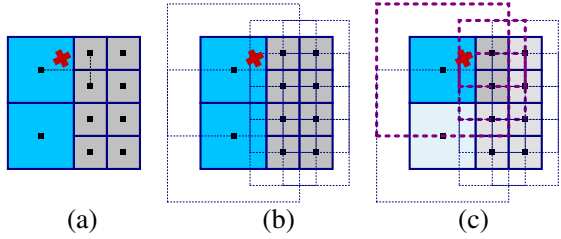}};
\node[anchor=south west,inner sep=0] at (0.8,-0.4) {(a)};
\node[anchor=south west,inner sep=0] at (3.4,-0.4) {(b)};
\node[anchor=south west,inner sep=0] at (6.0,-0.4) {(c)};
\end{tikzpicture}
\caption{\label{fig:dual}
Geometrical setup for reconstruction via basis functions as proposed in
related work by Wald et al.~\cite{wald:2017b}. (a) Cell-centric AMR data set with
boundary cells of neighboring levels. This presents us with the T-junction
problem. (b) Reconstruction of cell-centric data would \emph{locally} be
performed on the dual grids; neighboring dual grid cells form the \emph{domain}
(dotted squares) of the original grid cells that they overlap. (c) The domains
that overlap the sampling position determine the cells that contribute to the
linear combination from \cref{eq:basis}.
}
\end{figure}
A very common form is \emph{cell-centric} structured AMR as for example
produced by FLASH~\cite{Dubey2008}. A challenge of
cell-centric data is that the cell centers at level boundaries generally do not
line up along the principle axes---this is called the T-junction problem (see \cref{fig:dual})---and
consequently, first order interpolation cannot easily be mapped to the
customarily used tent reconstruction filter whose sample positions are aligned.
For those reasons, \removed{authors who used GPUs for volume rendering}\added{GPU-based AMR rendering codes either} concentrated
\removed{either} on vertex-centric data only, or did not perform interpolation at
all if the data was cell-centric\added{.}\removed{, as for example proposed by K{\"a}hler
et al. Their system used GPUs to perform volume
ray casting in a shader. The subgrids were first reorganized using a kd-tree
to obtain relatively large bricks of same-level cells, and the resulting bricks
were rendered using ordinary volume ray casting as it is also typically used
for structured volume data.}

\subsection{Real-Time Reconstruction}
Recent work on real-time AMR visualization has focused on high-quality
reconstruction with cell-centric data and at level
boundaries. \added{For an extended discussion we refer the reader to the
paper by Wald et al.}~\cite{wald:2020b}. Generally, if
the \removed{data is structured}\added{topology used
is a structured grid}, one can just use the dual grid to perform
reconstruction. In contrast to \removed{mere structured data}\added{that}, with AMR data the dual
grid cells generally do overlap, in which case
multiple cells affect the sample value at the sampling position
(\cref{fig:dual}b).

The basis function reconstruction method by Wald et al.~\cite{wald:2017b} \removed{computes the
following weighted linear combination to reconstruct sample values:}
\added{reconstructs the value using a weighted linear combination:}
\begin{equation}
B(p) = \frac{\sum_{C_i}
\Test{H_{C_i}}(p)C_{v_i}}{\sum_{C_i}\Test{H_{C_i}}(p)}.
\label{eq:basis}
\end{equation}
$\Test{H_{C_i}}$ is a tent-shaped basis function:
\begin{equation}
  \label{eq:hat}
    \Test{H_C}(p) = \prod_{\mathbf{d} \in x,y,z}\max{\bigg(1-\bigg(\frac{|C_{p\mathbf{d}}-p_{\mathbf{d}}|}{C_{w}}\bigg),0\bigg)},
  \end{equation}
where $C_w \in \mathbb{R}$ and $C_p \in \mathbb{R}^3$ denote the size and position of cell $C_i$
and $p \in \mathbb{R}^3$ refers to the sample point. \added{Locally,
using tent basis functions results in linear interpolation. But as the cell sizes at level
boundaries differ, so do the extents of the tent shapes. To correct
for coarser cells being over-represented, the linear combination is
hence weighted by the cumulated tent basis functions.}

\removed{Wang et al. introduced the octant reconstruction method,
which further decomposes the AMR cells into octants and performs interpolation
within these octants. The octant vertex values are computed to ensure a continuous
interpolant across level boundaries by interpolating values from the coarser
neighboring cells. Wang et al. improved upon the octant method
with generalized trilinear interpolation. Generalized trilinear interpolation
computes the octant vertex values at level boundaries by performing interpolation
within a polygon formed by the cell centers of the coarser cells. The improved
accuracy of the octant vertex values reduces artifacts in the reconstructed field,
improving the quality of isosurface rendering in particular.
Wang et al. demonstrated that a fixed number of polygon configurations exist, whose weights
can be precomputed to improve performance.}



\section{ExaBricks Data Structure}
\begin{figure*}[tb]
\centering
\begin{tikzpicture}
\node[anchor=south west,inner sep=0] (image) at (0,0) {\includegraphics[width=0.99\linewidth]{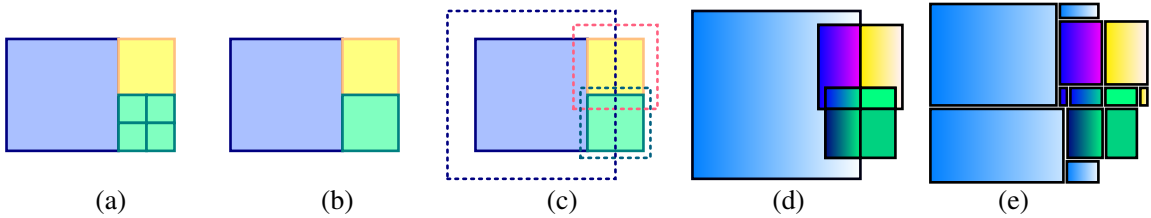}};
\node[anchor=south west,inner sep=0] at (1.2,-0.4) {(a)};
\node[anchor=south west,inner sep=0] at (4.2,-0.4) {(b)};
\node[anchor=south west,inner sep=0] at (7.2,-0.4) {(c)};
\node[anchor=south west,inner sep=0] at (10.2,-0.4) {(d)};
\node[anchor=south west,inner sep=0] at (13.2,-0.4) {(e)};
\end{tikzpicture}
\caption{\label{fig:abr} ExaBricks data structure proposed in prior work~\cite{wald:2020b} that our method builds
upon. (a) Exemplary AMR data set comprised of three levels and six subgrids. (b) We first
combine same-level subgrids to single grids (bricks) using a spatial hierarchy builder.
(c) For correct interpolation at level boundaries, we do not only have to integrate over
the bricks themselves, but over the bricks' domains which in general overlap. (d) The
regions \emph{where} the domains overlap form a space decomposition of generally concave
shapes. Those shapes we call the \emph{active brick regions}\removed{ (ABR)}. Each \removed{ABR}\added{such region} stores a
list of pointers to its respective bricks. (e) We decompose the \removed{ABRs}\added{active brick regions} into convex blocks
using yet another spatial decomposition, an example is shown in the illustration.
We build an OptiX BVH over those blocks. The BVH can be used to
perform
ray marching with adaptive sampling and space skipping, or point containment queries as proposed in this paper.
}
\end{figure*}
We base our particle tracing code off of the ExaBricks data
structure and visualization software~\cite{wald:2020b}, which is optimized for interactive rendering of AMR
data sets on GPUs equipped with ray tracing hardware.
ExaBricks added to the state of the art by enabling reconstruction with a first
order interpolant and adaptive sampling for ray marching-style algorithms, all without
the need to perform per sample cell location via costly tree traversal.
\removed{Similar to the approach
by K{\"a}hler et al.,}\added{When building the data structure that allows for fast AMR cell location on GPUs,} the ExaBricks software first drops the original
AMR hierarchy that is generated by the simulation code, then builds a spatial
subdivision of same-level cells to obtain bricks.
At this point,
where \removed{the approach by K{\"a}hler et al.}\added{traditional AMR visualization systems} would just render the bricks and
composite the intermediate results, the ExaBricks software constructs what
the authors call the \emph{active brick regions} \removed{(ABR)} (\cref{fig:abr}).

The \removed{ABRs}\added{active brick regions} are obtained by first computing all the bricks' domains (i.e.,
the region of space where at least one cell in a brick has non-zero
basis function weight (denominator term in.~\cref{eq:basis})).
The brick domains are computed by extending
the brick boundaries by an amount large enough to accommodate the interpolant
that is later used for reconstruction. The ExaBricks data structure uses the basis
function reconstruction method described above and hence requires domains whose bounds
extend the brick bounds by half a cell in each direction.
The size of these padding regions varies widely across different bricks
as it depends on the AMR level of the cells
contained inside. Consequently, domains of different bricks can overlap by almost arbitrary amounts (\cref{fig:abr}c), and any point in space could lie within an almost arbitrary number of domains of possibly different-level bricks.

Since any brick whose domain a point lies in will influence that point's basis function evaluation, evaluating at a given point requires finding and iterating over all the bricks whose domain the point overlaps. To quickly find all such bricks for a given point, \removed{the authors of}\added{Wald et al.}~\cite{wald:2020b} proposed to
\removed{compute what they called the Active Brick Regions (ABR)s: namely, the}\added{use the aforementioned active brick regions, which form a} decomposition of space into regions where all points from a given region overlap the same domains. Given the near-arbitrary overlap of the different domains,
the resulting same-domain regions are not generally rectangular any more; but with each domain being rectilinear they are also not arbitrarily shaped either, generally forming what are the 3D equivalents of L-shapes, T-shapes, etc.\ (see, e.g., the yellow box in \cref{fig:abr}d). 

Given these spatial regions, \removed{the authors of}\added{Wald et al.}~\cite{wald:2020b} further subdivided these L- and T-like shapes into rectilinear 3D boxes (using a spatial kd-tree like subdivision), and for each such box stored a list containing the references to all the bricks whose domains are active for the space covered by this box; as well as the level
of the finest cells contained by either of those bricks. The level will later
be used to determine an appropriate step size for adaptive sampling.

The resulting rectilinear boxes of this ExaBricks data structure are then easily amenable to be rendered on a GPU with OptiX---the\removed{ ABR} boxes \added{induced by the active brick regions} can directly
be used as user-defined primitives in a user
geometry BVH (\cref{fig:abr}e), and since these boxes do not overlap, any point in 3D space will only ever be contained in exactly one such primitive.

ExaBricks uses a ray generation program to implement both a
volumetric and an implicit isosurface ray caster. The ray casters step from \removed{ABR box to ABR box}\added{brick region to brick region}
by using (regular) OptiX intersections with primary rays; when processing an \removed{ABR}\added{active brick region}, the interval
$\left[t_{near},t_{far}\right]$ resulting from those intersections is integrated over using adaptive
sampling based on the finest cell size of the bricks the \removed{ABR}\added{active brick region} is derived from.

During ray marching, 
the basis function reconstruction method is used to reconstruct sample values
(the choice of interpolant is however not restricted to the basis method, other interpolants are supported but potentially
require larger domains and thus different \removed{ABR configurations}\added{configurations of active brick regions}).
In contrast to Wald et al.~\cite{wald:2017b},
ExaBricks does not need to perform cell location per sample position using
kd-tree traversal.
Instead, queries can just refer to the information that is stored by the
current \removed{ABR}\added{brick region}. In particular, the brick IDs stored by each \removed{ABR}\added{active brick region} allow the software
to perform sample location by just iterating over the \removed{ABR's}\added{regions'} bricks, and then locating
the sample value using a simple offset calculation. The resulting memory access patterns
are particularly well-suited for GPUs.

\section{Particle Tracing Using RT Core Point Containment Queries}
In the following we describe how we extend the ExaBricks framework to compute and later render
streamlines from \removed{flowfields}\added{vector fields} using a particle tracer. We accomplish this by combining the various
methods and data structures described above. More important than the individual techniques is
how we the integrate them in the presence of a ray tracing framework for scientific visualization.

\subsection{Integration Methods}
Approximation of integrals of ordinary differential equations (ODEs) can be achieved by Euler's method or more sophisticated methods, like Runge-Kutta integration. Euler's method estimates next particle positions by adding a scaled tangent vector to the current position (starting at the seed position), resulting in a new position and repeating this procedure until a 
termination criterion is reached.
While this is perfectly fine for linear functions, for more complex functions this leads to errors, depending on the vector length (step size). A better approach to approximately solving ODEs is the class of methods introduced by Runge and Kutta. The basic idea of these methods is to determine the step direction not only by the tangent at the current position but taking a weighted average of the value of tangents at midpoints. When averaging over four slopes, this results in the well known fourth order Runge-Kutta method. See the technical report by Ken Joy~\cite{joy:1999} for a derivation of those standard integration methods.

\subsection{Seed Points and Buffer Allocation}
It is the user's responsibility to generate the seed points to initialize the particle tracer, and to make sure that they fall within the bounds of the \removed{flowfield}\added{vector fields}.
Before the tracer is initiated, we also require the user to set the maximum number of timesteps.

With that information, we
initialize a buffer of size $\# seeds \times \# steps$ in GPU memory that can hold all the sample positions
and that is available to the OptiX ray tracing pipeline.
While this may seem excessive at first glance, we point out that the typical objective with particle tracers
is to generate relatively few but informative streamlines, as this technique is generally prone to
visual clutter.


\subsection{Progressive Particle Tracing}
Now that the seed points are available on the device, there are several different ways
that we could initialize our particle tracer; a natural
choice would be to initiate the tracer from within a separate ray generation program than the one
that executes the other visualization algorithms, and asynchronously overlap the execution
of the two ray generation programs. In practice however, we found it more convenient
to initialize and subsequently update the particle tracer from within the same ray generation program
that executes the other visualization algorithms.
As one usually uses a relatively small number of seeds---using hundreds or
even thousands of seeds usually leads to severe visual clutter---it is feasible to advance the
particles by a few time steps in parallel without the visualization becoming unresponsive.

We thus decided to use a progressive particle tracing approach where, before we render a frame, we
advance the tracer by a constant and small number of timesteps, devoting one GPU thread to each
particle. After that, the ray generation program will perform volume and surface rendering---the latter
step will also render the traces that were computed so far (but excluding the timesteps that were
generated during this ray generation launch, as those are not
present in the surface BVH yet).

\subsection{Particle Advection}
\begin{figure}
\centering
\begin{tikzpicture}
\node[anchor=south west,inner sep=0] (image) at (0,0) {\includegraphics[width=0.99\linewidth]{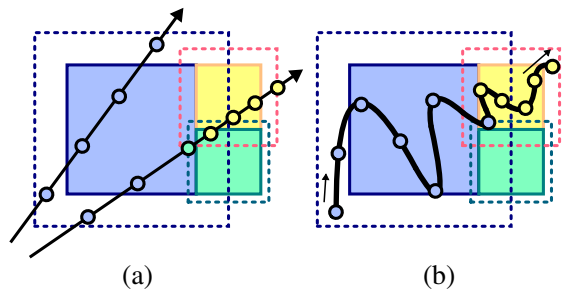}};
\node[anchor=south west,inner sep=0] at (1.5,-0.4) {(a)};
\node[anchor=south west,inner sep=0] at (5.5,-0.4) {(b)};
\end{tikzpicture}
\caption{\label{fig:adaptive} Adaptive sampling with a volume ray marcher (a) and with a particle tracer (b).
As the volume rays are a priori clipped against the active brick region boundary, it
is possible to determine exact step sizes for every sampling point. With particle tracing,
some sample points (like the blue sample point to the right) might use step sizes that do not match
the cell size in the current region.
}
\end{figure}
We use the ExaBricks data structure to realize parallel particle tracing for \removed{flow}\added{vector field} visualization
and implement that using RTX point containment queries. Using ExaBricks and its spatial decomposition
into \removed{ABRs}\added{active brick regions} has the major advantage that we can make use of hardware-accelerated cell location and that
the \removed{ABRs}\added{regions} enable adaptive sampling.

\begin{figure}
\centering







\includegraphics[width=0.98\columnwidth]{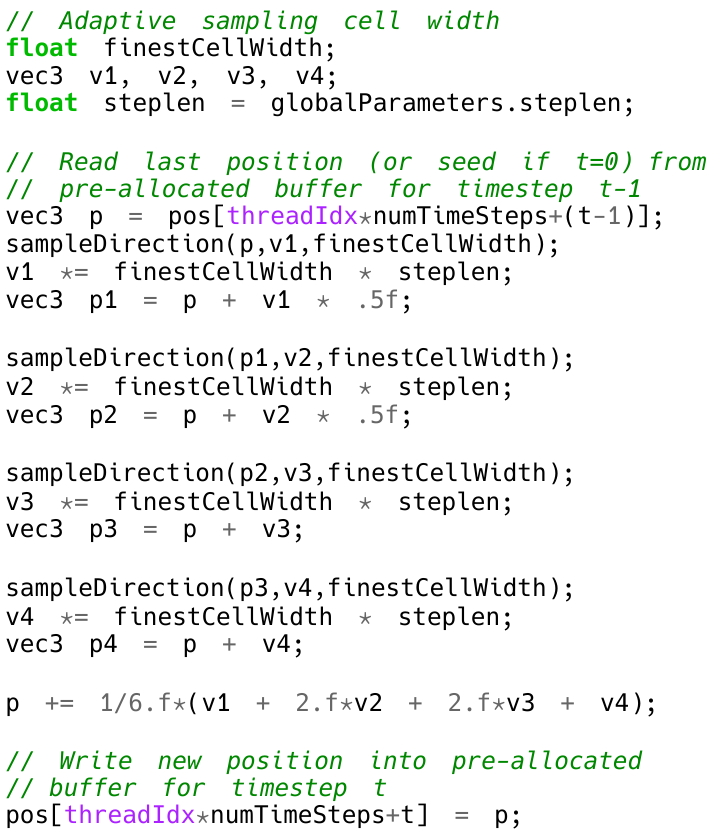}\\
\caption{\label{lst:rk-integration-step}\added{Performing a Runge-Kutta integration step.}
}
\end{figure}

Adaptive sampling is guided by the finest-level cell sizes stored with the \removed{ABRs}\added{brick regions}.
The adaptive sampling process is illustrated in \cref{fig:adaptive}.
A Runge-Kutta integration step is presented in \cref{lst:rk-integration-step}.

To initiate the integration step, each GPU thread is
responsible for one particle and will first retrieve its \emph{previous}
sample position (or, if \texttt{t=0}, its seed point) from the device buffer
(code handling particles whose traces terminated prematurely is omitted here for simplicity).

At the core of the algorithm is the function \texttt{sampleDirection()} that will
construct a direction vector using three point containment queries at the
3D sample position \texttt{p}
into the (user-configurable) $X,Y,Z$ channels associated with
the \removed{flow field}\added{vector field} (e.g., the three vector components of the velocity or vorticity field).
The \texttt{sampleDirection()} function also returns the finest cell width in the sampled
region using a modifiable C++ reference.

The \texttt{sampleDirection()} function performs basis method reconstruction
\cite{wald:2017b} by tracing one ray with length zero into the \removed{ABR}\added{region} BVH. Since \removed{ABR}\added{region} boxes do not overlap, the first primitive that contains the ray origin is exactly the \removed{ABR} box that lists all bricks that influence this sample point. Thus, all the intersection program has to do is check if the ray origin is inside the \removed{ABR}\added{active brick region's} bounds (it usually will be, but there is no guarantee that the BVH traversal might not call a false positive---so needs to be checked); then upon the first intersection store the pointer to the brick list in the per-ray-data for the \texttt{sampleDirection()} to iterate over, and terminate the ray. The \texttt{sampleDirection()} function then iterates over the bricks in the
list and computes the basis function values and weights for the $X,Y,Z$ channels of
the \removed{flowfield}\added{vector field}. The reconstructed 3D vector serves as one of the four direction vectors that are required by the fourth order Runge-Kutta integration method (\texttt{v1} through \texttt{v4} in the code listing above).

The four Runge-Kutta direction vectors returned by \texttt{sampleDirection()}
are \emph{individually} scaled not only by the constant and user-configurable
step length, but also by the finest cell width in the \removed{ABR}\added{brick region}. That way we adaptively sample the \removed{flowfield}\added{vector field}.
Finally, the newly computed particle position
is written back into the sample positions buffer, at the position that is reserved for the 
current timestep.

In contrast to volume ray marching, with the particle tracer it is not always possible to
choose the best step size for adaptive sampling, as can be seen in
\cref{fig:adaptive}b. Volume rays are first clipped against the \removed{ABR's}\added{active brick region's} boundary, so that we
know the exact length of the ray segment covered by the \removed{ABR}\added{region}. With particle tracing, when
we choose an adaptive sampling step size, we cannot easily determine a priori if the curve
will bend into a brick region with finer cell sizes and thus potentially undersample the
volume at level boundaries. In practice, and with the data sets we used for our tests,
we did not find this to cause any noticeable issues, but still want to point out this
potential source of undersampling inherent to the approach.

\subsection{High-Quality Rendering with Curved Geometries}

\begin{figure}[tb]
\centering
\includegraphics[width=0.999\linewidth]{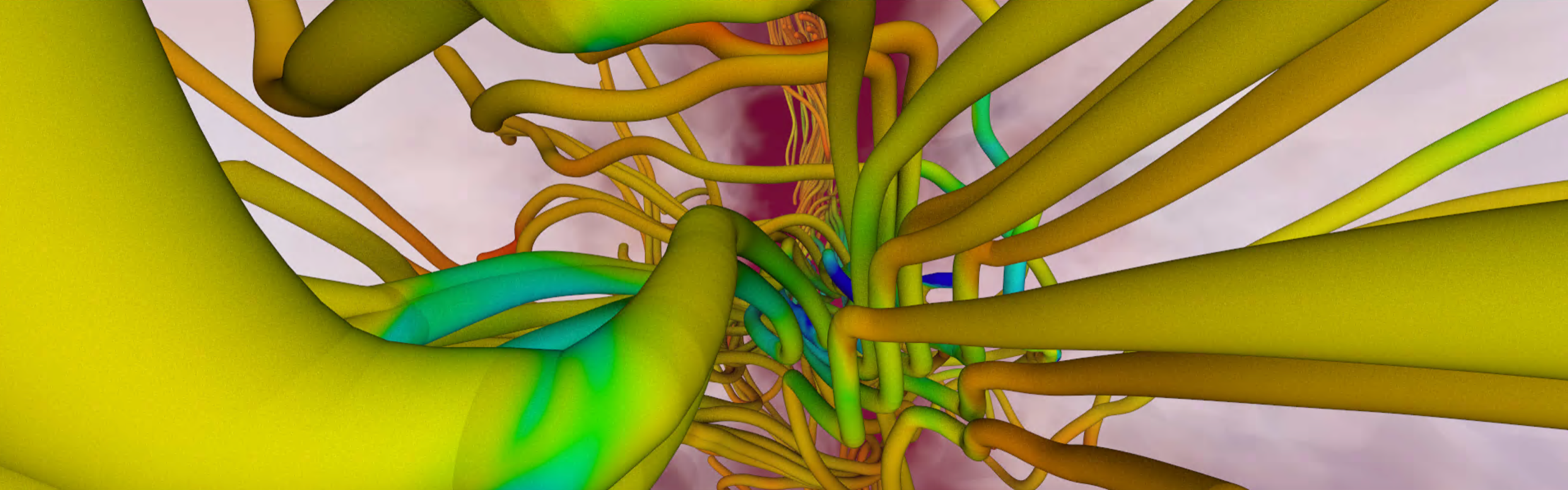}
\caption{\label{fig:tubes} Zoomed-in \removed{flowfield}\added{vector field} visualization. Integration with a
ray tracing framework enables high-quality visualization compared to the omnipresent
simple 2D line strips that many rasterization-based visualization systems use to represent streamlines.
}

\end{figure}
At this point, when all the timesteps were computed, we are finished with the actual 
post-processing and particle advection procedure. With the sample positions buffer
filled, we can now proceed to render the traces, for example
as streamlines.

How this is done is largely independent of the actual particle advection step.
As we are using OptiX~7, one option would be to use the integrated curve primitives
to visualize the streamlines. For technical reasons, our system however uses the
tube primitives presented by Wald et al.~\cite{wald:2020c} and just renders
the control points as rounded cylinder strips (\cref{fig:tubes}). The BVH for the geometry is rebuilt whenever
the tracer, in a previous ray tracing pipeline launch, generated new sample points.

Integration with a ray tracing-based pipeline of course also enables high-quality rendering
and shading techniques (also \cref{fig:tubes}).
ExaBricks implements a simple local shading model that is augmented with ray-traced ambient
occlusion. It would however be straightforward to also support more advanced rendering 
algorithms. Optionally, we can also map an AMR field (e.g., velocity magnitude) as a color onto the streamlines.

\section{Example Applications from Astrophysics}
We present two user case studies where our system was used for post-processing followed
by visual exploration to gain insights into the astrophysical data sets. The two
data sets represent phenomena at different scale and are both time-varying.
They illustrate the relevance of AMR \removed{flowfield}\added{vector field} exploration using high-quality 
rendering for this particular scientific field.


\subsection{Example Application 1: Magnetic Fields in Star-Forming Molecular Clouds}
\label{sec:cloud}

\begin{figure*}[tb]
\centering
\begin{tikzpicture}
\node[anchor=south west,inner sep=0] (image) at (0,0) {\includegraphics[width=0.999\linewidth]{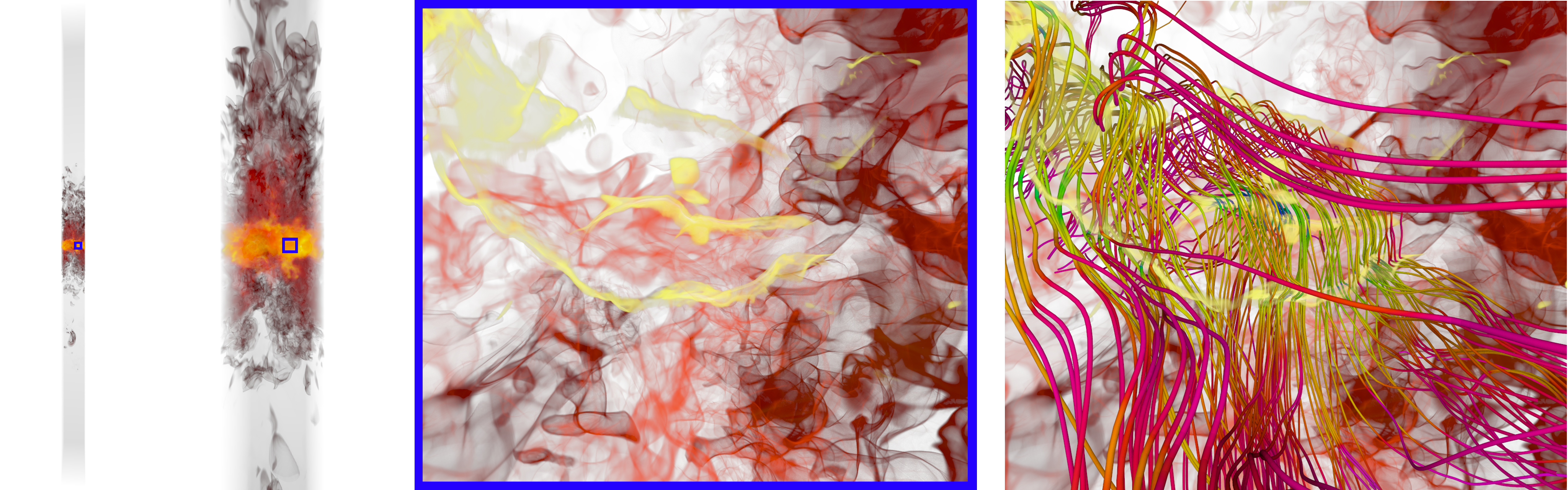}};
\node[anchor=south west,inner sep=0] at (0.53,-0.4) {(a)};
\node[anchor=south west,inner sep=0] at (2.45,-0.4) {(b)};
\node[anchor=south west,inner sep=0] at (6.6,-0.4) {(c)};
\node[anchor=south west,inner sep=0] at (12.4,-0.4) {(d)};
\end{tikzpicture}
\caption{\label{fig:silcc} Graphical representation of the SILCC-Zoom simulation zooming in from the galactic scale (a, b) onto the molecular cloud (c) and its embedded filamentary substructure (yellow gas in panels c and d). Panel (d) shows the complex magnetic field structure  associated with the molecular cloud, and reveals the dependencies between the filaments and the magnetic flow.
}
\end{figure*}
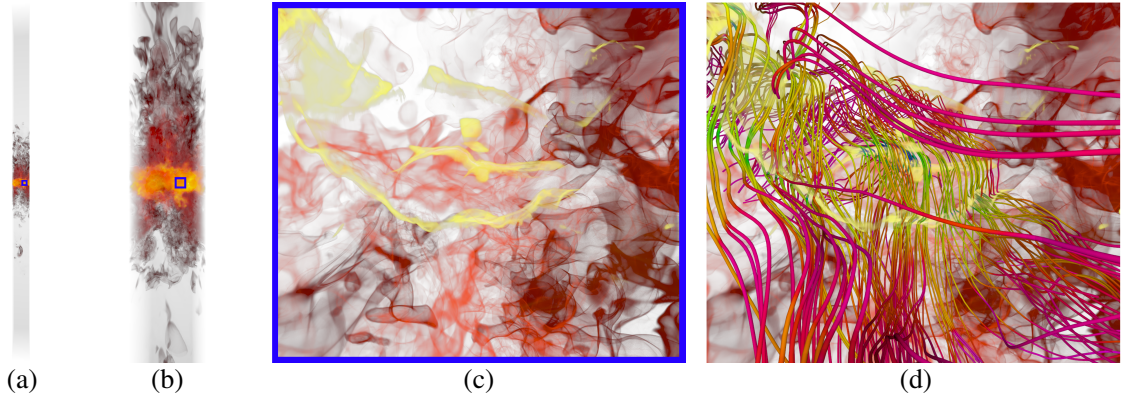

One key research question in modern astrophysics is how stars in our Galaxy (the Milky Way) are forming.
Broadly speaking, stars condense out from the diffuse gas sitting between the already existing stars in our Galaxy. Due to the effect of gravity, large (several ten light-years in diameter) accumulations of this gas form so-called molecular clouds, which represent the nurseries of forming stars. As time evolves, parts of these molecular clouds become denser and denser due to the effect of gravity, ultimately leading to the formation of a new generation of stars.

Hence, understanding these properties is crucial for understanding the formation of individual stars. From observations it is known that molecular clouds have a highly complex and possibly hierarchical structure \cite{Roman2010}. In addition, many physical processes like turbulence, gravity, radiation and magnetic fields influence the evolution of molecular clouds in a complex interplay.
This in turn necessitates dedicated 3D simulations to model the formation and evolution of molecular clouds---and their embedded stars.

In the SILCC-Zoom project \cite{Seifried2017} the evolution of molecular clouds, which are embedded in a part of a spiral galaxy (in turn modelled in the larger-scale SILCC project, \cite{Walch2015}), is modelled by means of state-of-the-art, 3D, magneto-hydrodynamical simulations. For this purpose, the versatile astrophysical AMR code FLASH is used \cite{Dubey2008}, which was modified
to include the various physical processes mentioned before to allow for one of the most realistic molecular cloud simulations to date.

In \cref{fig:silcc} we visualize one snapshot of the simulation, gradually zooming in from the largest onto the smallest scales.
This particular snapshot consists of 72.8~M cells and 142~K AMR subgrids that are distributed across seven hierarchy levels. The logical grid---an imaginary
structured grid that, if the simulation was resampled on it, would
retain the detail of even the finest AMR level---has a size
of $4096 \times 4096 \times 81920$ cells.
This unveils an intrinsic complication of astrophysics that the characteristic length scales extend over a significant range often covering \emph{four orders of magnitude and more}, which explicitly requires the usage of AMR.
In the SILCC-Zoom simulation, the represented scales extend from galactic scales ($\geq$ 1 kpc, \cref{fig:silcc}a and \cref{fig:silcc}b), over molecular clouds (1 - 100 pc, \cref{fig:silcc}c and \cref{fig:silcc}d) down to their complex substructure (0.1 - 1 pc, yellow material in \cref{fig:silcc}c and \cref{fig:silcc}d). This substructure consists of so-called \emph{filaments}, i.e.\ elongated dense structures embedded in gas of lower density, which arrange in a complex network. Filaments are very thin compared to the entire molecular cloud, i.e.\ have diameters of less than one lightyear, corresponding to 1\% of the cloud’s diameter. Hence, to resolve this filamentary structure, high resolution facilitated by AMR is required.

Despite their small size, filaments are key to understanding the star formation process:
They present the densest structures of the molecular cloud and thus stars will form preferentially inside them. 


%

%
One way
to assess the importance of magnetic fields is their orientation relative to the dense gas, i.e.\ the filaments. 
Observational and theoretical studies predict that magnetic fields pierce perpendicularly through the center of filaments and can be dragged along with the filament. As an analogy one could imagine a pearl on a rubber band: pulling the pearl to the side will bend the rubber band, but it will still pierce through the pearl’s center.



Gaseous phenomena that can be expressed as scalars, like the density field, can
be effectively rendered using the typical volume and isosurface rendering modalities. In
\cref{fig:silcc}c the filaments for example are the structures that are
assigned the color yellow and can be easily differentiated from the
surrounding gas via an RGBA transfer function. For 3D vector fields such assignments are,
however, often ineffective: the vector components can easily be mapped to RGB components
and rendered as a volume,
but such a visualization will not convey directionality.
Using a \removed{flowfield}\added{vector field} visualization like the one proposed, with high-quality
ray tracing and ambient occlusion, the spatial relationship is however revealed. As
can be nicely seen in \cref{fig:silcc}d, the \removed{streamlines}\added{fieldlines} associated with the magnetic
field pierce through and bend around the dense, gaseous filaments.

A challenge unrelated to the point containment query technique, but inherent to \removed{flowfield}\added{vector field}
visualization, that we encountered, was the occurrence of visual clutter
when not carefully choosing seed points. We therefore, in a pre-process, isolated the
filaments by identifying the AMR cells with the highest density, and placed random seeds
by choosing from the set of cells that belong to the filaments. It is noteworthy here that
we found the integration of the technique into a ray tracing framework beneficial as we can
easily make use of techniques such as ambient occlusion, which can help to reduce visual
clutter and provide additional depth cues for the visualization.


\subsection{Example Application 2: Binary Neutron Star Formation via Common-envelope Ejection}

\begin{figure*}[tb]
\centering
\includegraphics[width=0.24\linewidth]{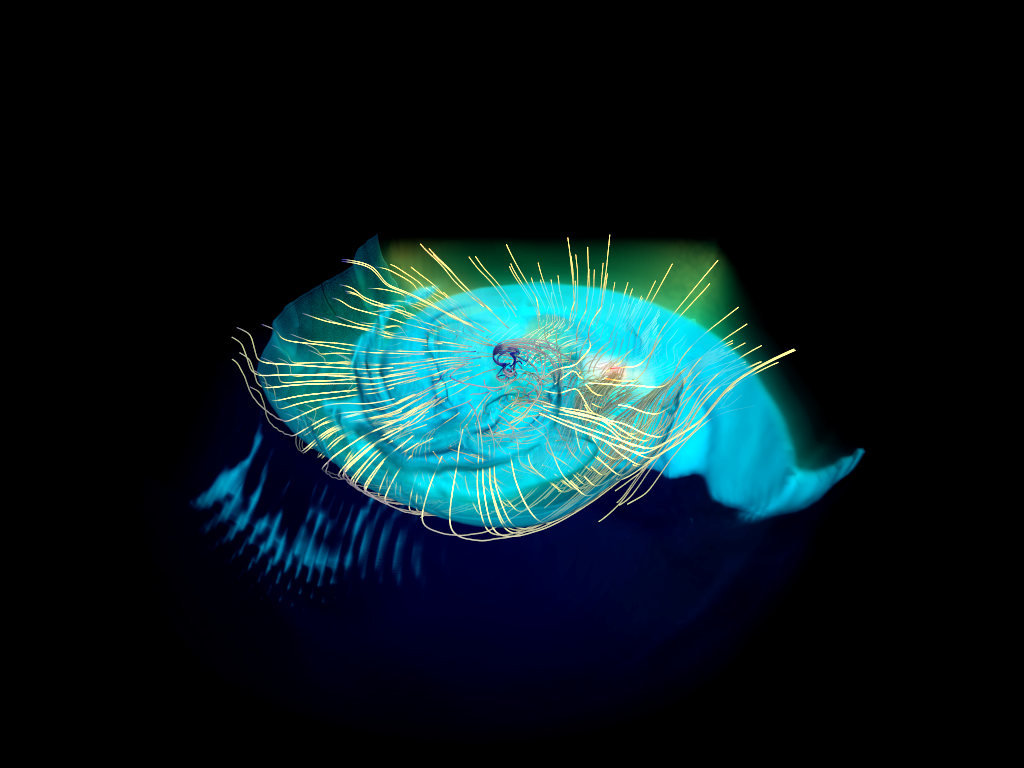}
\includegraphics[width=0.24\linewidth]{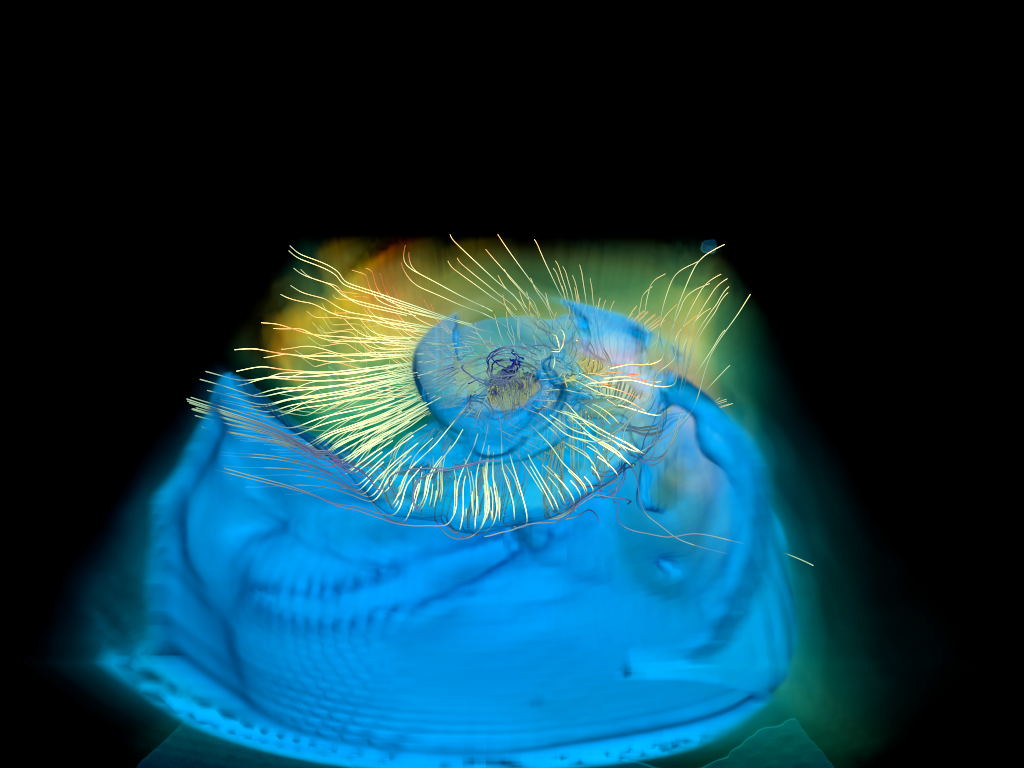}
\includegraphics[width=0.24\linewidth]{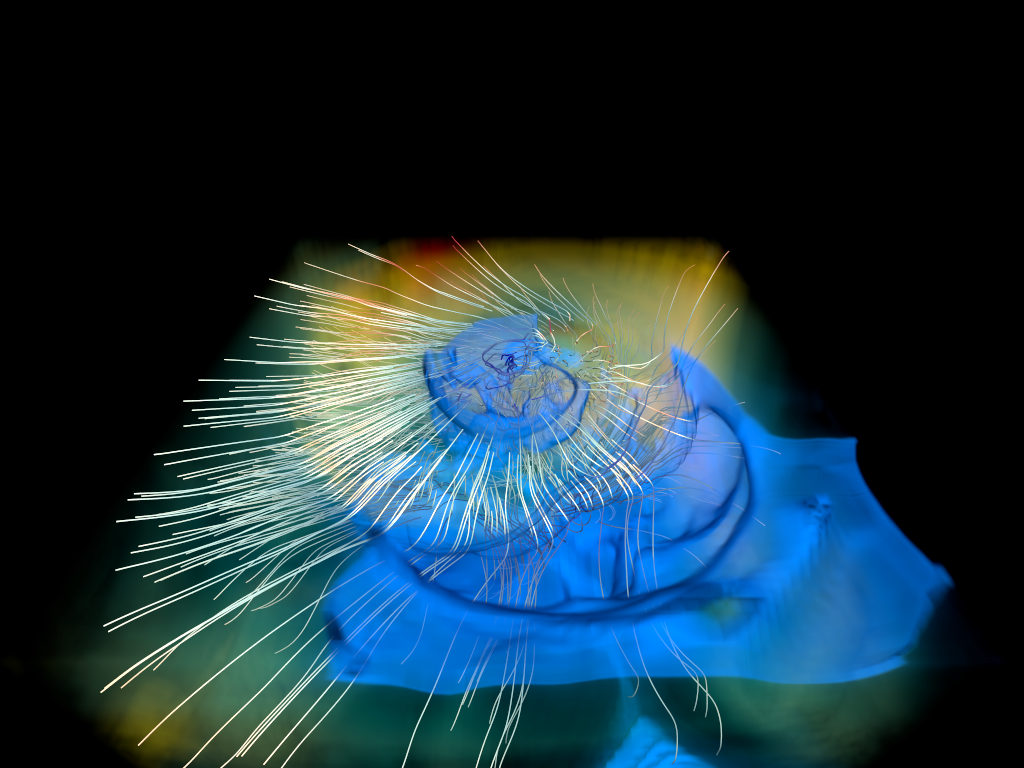}
\includegraphics[width=0.24\linewidth]{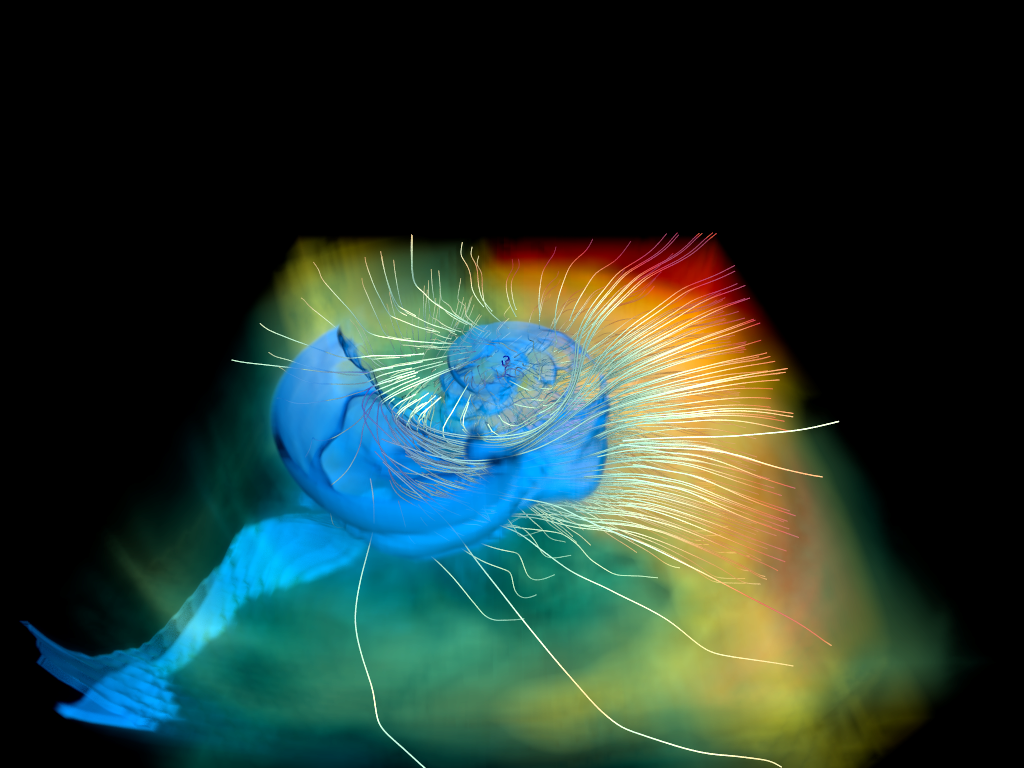}\\
\vspace{1em}
\includegraphics[width=0.24\linewidth]{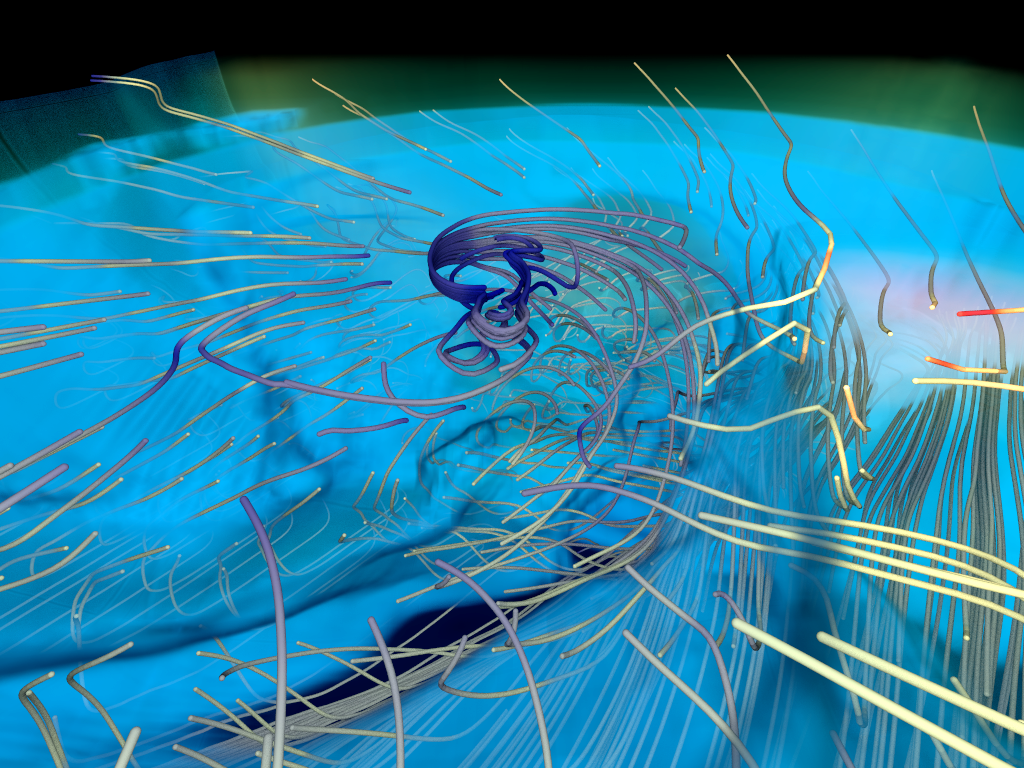}
\includegraphics[width=0.24\linewidth]{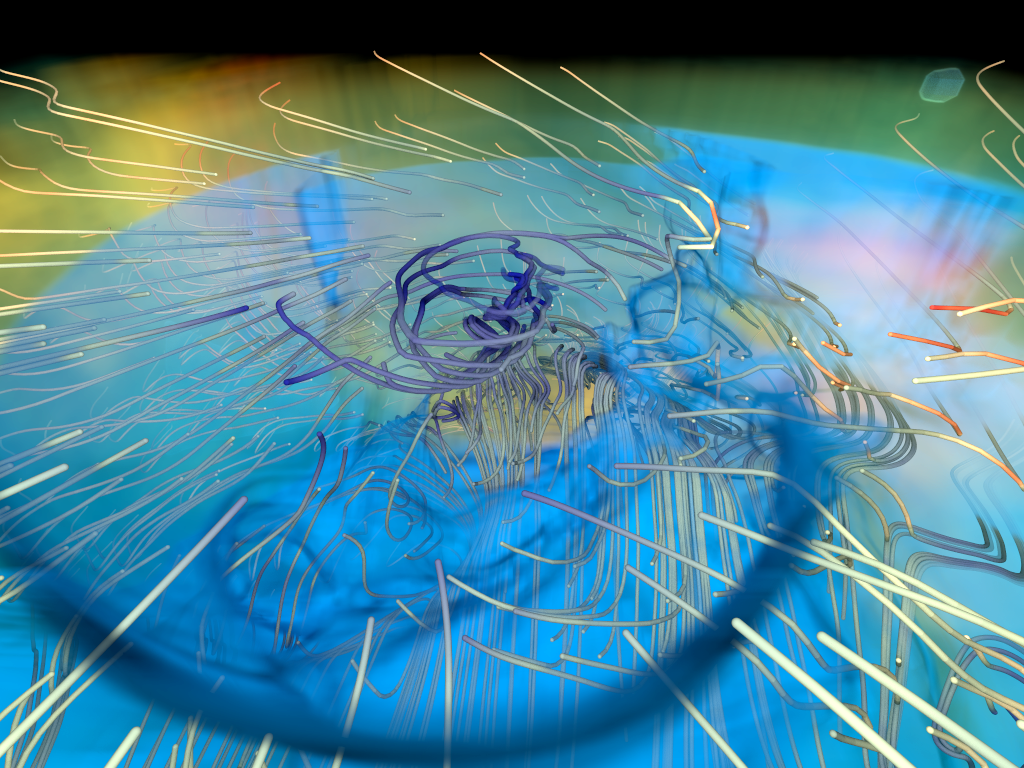}
\includegraphics[width=0.24\linewidth]{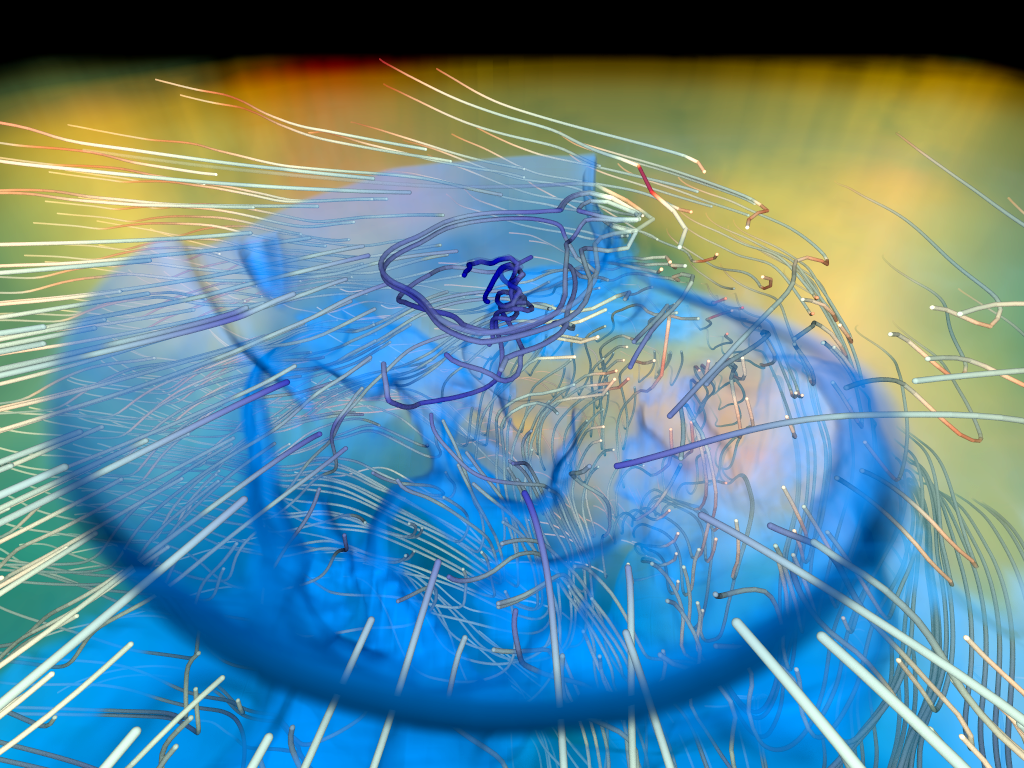}
\includegraphics[width=0.24\linewidth]{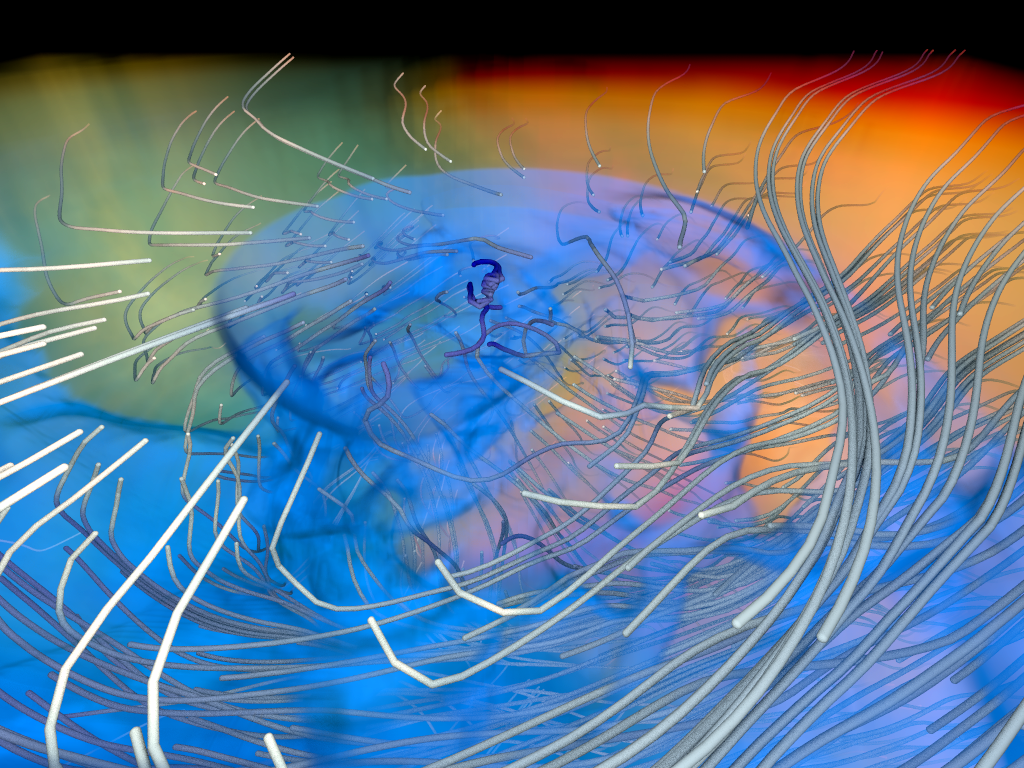}
\caption{
\label{fig:p189} 
Binary neutron star formation via common-envelope ejection.
3D rendering of the ratio of velocity magnitude to local escape velocity with streamlines as a neutron star inspirals through the envelope of a gas giant.
Time evolves from left to right. Bottom panels are same time steps as top panels, but zoomed in.
Gas is ejected from the initially gravitationally-bound envelope of the gas giant, allowing the neutron star to reach a final orbital separation before the gas giant evolves to a neutron star. The two neutron stars will then merge via emission of gravitational radiation (gravitational waves).
}
\end{figure*}




In a landmark discovery, astronomers have observed the merger of two neutron stars in both gravitational waves and electromagnetic radiation (from radio to gamma-rays).
This discovery has opened up new lines of research in several areas of  physics and astrophysics, from understanding gravity in the strong-field regime to understanding nucleosynthesis in the universe.

However, we do not understand how these systems are formed. In order to merge via emission of gravitational radiation within the age of the universe, two neutron stars must have a relatively small orbital separation, much smaller than the initial separations or even radii of their progenitor stars (the red giants that exploded in supernovae and left neutron stars behind). If the neutron stars started out at the orbital separations of their progenitor stars, they would not merge within the age of the universe.

One proposed solution to this problem is that the neutron stars tighten their orbital separation during the poorly-understood ``common-envelope'' phase. During this phase, one star (say, that has already evolved to a neutron star) may interact with the envelope of the other star (a red giant) and lose orbital energy, causing it to inspiral toward the center of the other star. This process will not stop, and the neutron star will merge with the core of the red giant,
unless the neutron star is able to eject the envelope of the red giant and ``park'' its orbit in order to reach a final orbital separation around the core of the red giant. The remaining core of the red giant will then explode and form a second neutron star. The two neutron stars will then lose energy to gravitational waves, causing their orbit to shrink over time. If they start close enough, they will merge within the age of the universe and be observable in both gravitational and electromagnetic radiation, a so-called ``multi-messenger'' event.

This process of common envelope ejection has only recently been studied in 3D hydrodynamics~\cite{law-smith_2020}. 
This study also used the AMR code FLASH,
but the physical scale of the star-forming molecular clouds described in the previous section is of order 1 kpc, whereas the physical scale of the binary neutron star system is of order 1000 solar radii, with the neutron star having a radius of 10 km.
The timescales involved are also quite different. For the star-forming molecular clouds described in the previous section, the relevant physical processes occur on timescales of order $10^6$ years, whereas the common envelope ejection occurs on timescales of order $1$ day. 
Thus, both the spatial and temporal resolution of the two data sets is quite different.

In this study, a neutron star orbits the core of a red supergiant and successfully ejects its envelope after a few orbital passages. This allows the neutron star to reach a final orbital separation before the red giant evolves to a neutron star, and thus for the eventual neutron star merger to be observable.
\cref{fig:p189} shows the passage of the neutron star through the envelope of the red giant. 

In the visualization, the streamlines track the velocity flow of the gas, which is imparted to it by the orbital motion of the neutron star.
There is a velocity flow both locally, surrounding the orbiting neutron star (this shows the flow of material around the neutron star and the possibility of Bondi-Hoyle-Lyttleton accretion of material onto the neutron star---this is potentially observable as a secondary electromagnetic source, and is also important for determining the growth of the neutron star during its orbit; if it accretes too much material, the neutron star will collapse to a black hole), and globally, as a result of multiple passages of the neutron star, showing the ejection of gas material in the common envelope.
The flow visualization conveys how the gas is shocked approximately radially outward after the neutron star sweeps it out. 

\section{\added{Performance Measurements}}
\begin{figure}[tb]
\centering
\includegraphics[width=\linewidth]{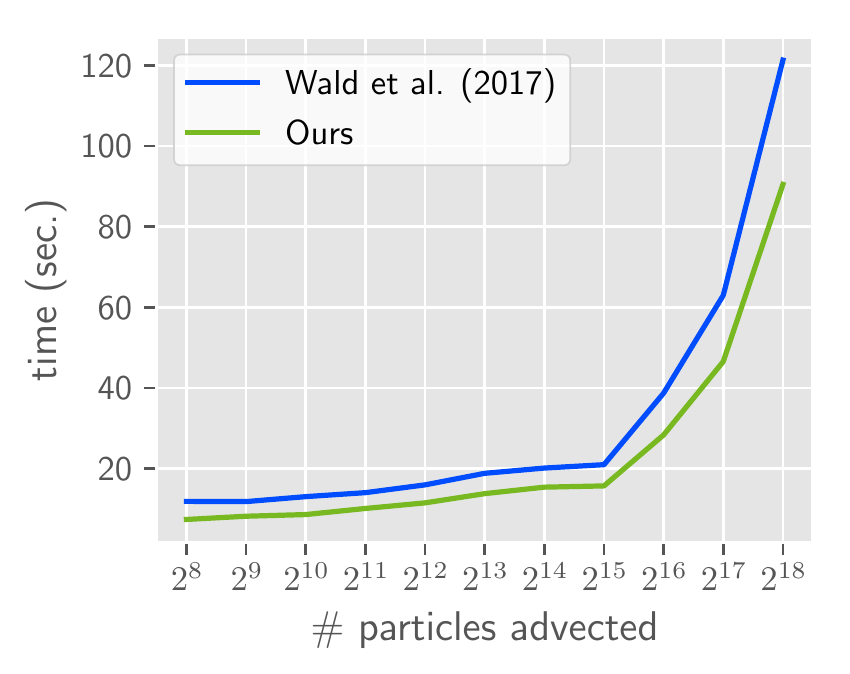}
\caption{\label{fig:results} 
\added{Execution times for advecting particles 100K times through the
magnetic field of the molecular cloud data set example application.}
}
\end{figure}
\added{
We present results of a comparative performance evaluation in \cref{fig:results}.
There we compare against the kd-tree method by Wald et al.~\cite{wald:2017b}.
We however built the kd-tree over the brick set induced by the ExaBricks data
structure~\cite{wald:2020b}. With particle advection, and as opposed to, e.g., direct
volume rendering, we cannot make use
of the optimization where the tree is only traversed once per active brick region.
The performance study is hence a direct comparison of software kd-tree traversal versus
hardware-accelerated OptiX BVH traversal, as the samples taken will be exactly the same.
For our comparison, we advect a varying number of particles and perform 100~K steps
per particle through the molecular cloud data set
from our first example application. We observe how, with increasing input sizes, the GPU becomes fully saturated at around
$2^{15}$ particles advected and we gradually benefit from hardware-accelerated tree traversal. The advantage of using ray
tracing cores becomes more pronounced the more particles we advect.}

\section{Discussion and Conclusion}
We presented an unconventional yet effective technique to perform point containment queries into AMR
vector fields. The technique is quite general and has previously also been used by other authors,
e.g., to accelerate point containment queries in finite element data sets and other
compute-intensive applications. We adapted the technique to AMR \removed{flowfield}\added{vector field} visualization but
note that the overall idea of mapping 3D tree traversal to a ray tracing problem and by that
enabling the use of RT cores is generally applicable to all sorts of problems.
Using ray tracing in this holistic way also potentially makes integration of high-quality 
rendering easier than with typical scientific visualization systems that use rasterization.
Conversely, as such a system usually maintains its 3D geometry in a BVH anyway, using that
for point containment queries is a convenient choice with very little overhead.


The major advantage of using RTX point containment queries, in our case, is that we can
leverage adaptive sampling when extracting traces from the \removed{flowfield}\added{vector field}, while other methods,
such as direct volume rendering, will potentially benefit more if execution of multiple threads
stays within the same active brick region for some time. Leveraging this effect with our approach
is not possible. In the future, we however plan to investigate if an approach that is not
based on ray tracing but instead parallelizes the particle advection step over the set of
\removed{ABRs}\added{active brick regions} can result in increased performance. Similar approaches based on macrocells from structured grids however often suffer from
scalability issues, and the adjacency information that is required to advect the particles from
one \removed{ABR}\added{region} to another is non-trivial to maintain, compared to the macrocell connectivity of a grid accelerator
built on top of a uniform grid topology.

\added{Future work will also include adding support for path lines 
and unsteady flow---the examples we presented were based on
streamlines and fieldlines extracted from single animation frames.
A major challenge is that AMR topologies often change
over time and hence the ExaBricks data structure needs expensive
rebuilds, so that addressing this problem would also require to address
interactive data structure construction.}

We argue that the approach of using RTX ray tracing cores for point containment queries is
so general that it can be used for all sorts of different applications, ranging from
physics like ours to engineering. \added{Such applications are not limited to
structured or semi-structured grids, but also to generally unstructured, finite elements. The method could also potentially be extended to support multiple vector
fields.}
In this respect, our paper is also meant to serve as an example and
potentially a guide for practitioners to implement this technique in their own frameworks.


\section*{Acknowledgments}

SZ acknowledges funding by the Deutsche Forschungsgemeinschaft (DFG, German Research Foundation)---grant no.~456842964.
SZ, DS, SW, and AH wish to thank the Ministry of Culture and Science of the State of North Rhine-Westphalia for supporting the work through the PROFILBILDUNG grant PROFILNRW-2020-038C.
DS and SW acknowledge funding by the Deutsche Forschungsgemeinschaft (DFG) via the Collaborative Research Center SFB 956 “Conditions and Impact of Star Formation” (subprojects C5 and C6). SW further acknowledges funding from the European Research Council via ERC Starting Grant no. 679852 (RADFEEDBACK). 
J.L.-S. acknowledges use of the $lux$ supercomputer at UCSC, funded by NSF MRI grant AST 1828315, and the HPC facility at the University of Copenhagen, funded by a grant from VILLUM FONDEN (project number 16599).
The SILCC-Zoom simulation data used in this work is based on the FLASH code, which was in part developed by the DOE NNSA-ASC OASCR Flash Center at the University of Chicago. The simulations have been run on Supermuc at the Leibniz-Rechenzentrum Garching using supercomputing time acquired via the Gauss Centre for Supercomputing.

\bibliographystyle{IEEEtran}
\bibliography{stefan,exabrick,daniel} 

\begin{IEEEbiography}{Stefan Zellmann}{\,}
received the graduate degree in information systems from the
University of Cologne and the doctor's degree in computer science in 2014, with a
PhD thesis on high performance computing and direct volume rendering.
His current affiliation is with the Institute of Visual Computing at the Bonn-Rhein-Sieg University of Applied Sciences.
He was with the Chair of Computer Science, University of Cologne from 2009 to 2021, where he still serves as a lecturer.
His research focuses on the interface between high performance computing and real-time rendering.
There he is primarily concerned with algorithms and software abstractions to leverage real-time ray tracing and photorealistic rendering at scale and for large 3D models and scientific data sets. Contact him at stefan.zellmann@h-brs.de
\end{IEEEbiography}

\begin{IEEEbiography}{Daniel Seifried}{\,}is a PostDoc in the field of astrophysics at the I. Physical Institute at the University of Cologne, Germany. His research interests include 3D, magneto-hydrodynamical simulations of star formation. Contact him at seifried@ph1.uni-koeln.de 
\end{IEEEbiography}

\begin{IEEEbiography}{Nate Morrical}{\,}
is a PhD student at the University of Utah, an intern alumni from Nvidia and Pixar, and is currently working under Valerio Pascucci as a member of the CEDMAV group in the Scientific Computing and Imaging Institute (SCI). His research interests include high performance GPU 
computing, real time ray tracing, and human computer interaction. Prior to joining SCI, Nate received his B.S. in Computer Science from Idaho State University, where he researched interactive computer graphics and computational geometry under Dr. John Edwards. Email: natemorrical@gmail.com
\end{IEEEbiography}

\begin{IEEEbiography}{Ingo Wald}{\,} is a director of ray tracing at NVIDIA. He received his master’s degree from Kaiserslautern University and his PhD from Saarland University, then served as a postdoctorate at the Max-Planck Institute Saarbrücken, as a research professor at the University of Utah, and as technical lead for Intel’s software-defined rendering activities (in particular, Embree and OSPRay). Ingo has coauthored more than 90 papers, multiple patents, and several widely used software projects around ray tracing. Email: iwald@nvidia.com
\end{IEEEbiography}

\begin{IEEEbiography}{Will Usher}{\,} is a Ray Tracing Software Engineer at Intel, working on the oneAPI Rendering Toolkit. His work focuses on CPU and GPU ray tracing, distributed rendering, and scientific visualization. He completed his Ph.D. in Computer Science in 2021 on data management and visualization for massive scientific data sets at the Scientific Computing and Imaging Institute at the University of Utah, advised by Valerio Pascucci. Email: will.usher@intel.com
\end{IEEEbiography}

\begin{IEEEbiography}{Jamie A.P. Law-Smith}{\,} 
is a fellow at the Institute for Theory and Computation at the Center for Astrophysics $|$ Harvard \& Smithsonian. 
His research is in theoretical astrophysics and physics; some interests include tidal disruptions of stars by black holes, common envelope evolution in the context of gravitational-wave sources, and de Sitter space in theories of quantum gravity.
He received his Ph.D. in Astronomy and Astrophysics from the University of California Santa Cruz. 
Email: jamie.law-smith@cfa.harvard.edu
\end{IEEEbiography}

\begin{IEEEbiography}{Stefanie Walch-Gassner}{\,}
is a Full Professor of theoretical astrophysics leading the theoretical astrophysics group and is the managing director of the I. Physics Institute of the University of Cologne, Germany. She is also vice director of the Center for Data and Simulation Science (CDS) at the University of Cologne. Her research interests include the physics of the interstellar medium, star formation and stellar feedback as well as the development of new methods and algorithms for high-performance computing, in particular magneto-hydrodynamics, radiative transfer, and astro-chemistry. Email: walch@ph1.uni-koeln.de
\end{IEEEbiography}

\begin{IEEEbiography}{Andr\'{e} Hinkenjann}{\,} is a Full Professor of computer graphics and interactive environments with the Department of Computer Science, Bonn-Rhein-Sieg University of Applied Sciences (BRSU), Sankt Augustin, Germany, and an Adjunct Professor with the Department of Computer Science, University of New Brunswick, Fredericton, NB, Canada. He is the Founding Director of the Institute of Visual Computing, BRSU. His research interests include high-quality rendering, interactive data exploration, and high-resolution display systems., Prof. Hinkenjann is a member of ACM SIGGRAPH, Eurographics, IEEE and German GI. He is a regular reviewer for international conferences and journals on the above mentioned topics. Email: andre.hinkenjann@h-brs.de
\end{IEEEbiography}

\end{document}